\documentclass[onecolumn,prd,superscriptaddress,showpacs,floatfix,%
preprintnumbers, nofootinbib,hyperref]{revtex4}  
\usepackage{epsfig}   
\usepackage{bm}  
\usepackage{color}

\newcommand{\ben}{\begin{equation}}
\newcommand{\een}{\end{equation}}

\newcommand{\gtrsim}{\,\rlap{\lower3.7pt\hbox{$\mathchar\sim$}}
\raise1pt\hbox{$>$}\,}
\newcommand{\lesssim}{\,\rlap{\lower3.7pt\hbox{$\mathchar\sim$}}
\raise1pt\hbox{$<$}\,}

\newcommand{\be}{\begin{equation}}
\newcommand{\ee}{\end{equation}}  
\newcommand{\bea}{\begin{eqnarray}}
\newcommand{\eea}{\end{eqnarray}}

\def\theta{\vartheta}   

\begin{document}

\preprint{LAPTH-045/11}

\title{Probing the neutrino mass hierarchy with the rise time of a supernova burst}

\author{Pasquale D.~Serpico}
\affiliation{LAPTh, Univ. de Savoie, CNRS, B.P.110, Annecy-le-Vieux F-74941, France}

\author{Sovan Chakraborty}
\affiliation{II Institut f\"ur Theoretische Physik, Universit\"at Hamburg, Luruper Chaussee 149, 22761 Hamburg, Germany}

\author{Tobias Fischer}
\affiliation{GSI, Helmholtzzentrum f\"ur Schwerionenforschung GmbH,
Planckstr. 1, 64291 Darmstadt, Germany}
\affiliation{Technische Universit\"at Darmstadt, Schlossgartenstr. 9, 64289 Darmstadt,
Germany}

\author{Lorenz H\"udepohl}
\affiliation{Max-Planck-Institut f\"ur Astrophysik, Karl-Schwarzschild-Str. 1, 85748 Garching, Germany}

\author{Hans-Thomas Janka} 
\affiliation{Max-Planck-Institut f\"ur Astrophysik, Karl-Schwarzschild-Str. 1, 85748 Garching, Germany}

\author{Alessandro Mirizzi}
\affiliation{II Institut f\"ur Theoretische Physik, Universit\"at Hamburg, Luruper Chaussee 149, 22761 Hamburg, Germany}


\begin{abstract}
The rise time of a Galactic supernova (SN) $\overline{\nu}_e$ lightcurve, observable at a high-statistics experiment such as the  Icecube Cherenkov detector, can provide a diagnostic tool for the neutrino mass hierarchy at ``large'' 1-3 leptonic mixing angle $\theta_{13}$. Thanks to the combination of matter suppression of collective effects at early postbounce times on  one hand and the presence of the ordinary Mikheyev-Smirnov-Wolfenstein effect in the outer layers of the SN on the other  hand, a sufficiently fast rise time on ${\cal O}$(100) ms scale is indicative of an inverted mass hierarchy. We investigate results from an extensive set of stellar core-collapse simulations, providing a first exploration of the astrophysical robustness of these features. We find that for all the models analyzed (sharing the same weak interaction microphysics) the rise times for the same hierarchy are similar not only qualitatively, but also quantitatively, with the signals for the two classes of hierarchies significantly separated. We show via Monte Carlo simulations that the two cases should be distinguishable at IceCube for SNe at a typical Galactic distance 99\% of the times.  Finally, a preliminary survey seems to show that the faster rise time for inverted hierarchy as compared to normal hierarchy is a qualitatively robust feature predicted by several simulation groups. Since the viability of this signature ultimately depends on the quantitative assessment of theoretical/numerical uncertainties, our results motivate  an extensive campaign of comparison of different code predictions at early accretion times with implementation of  microphysics of comparable sophistication, including effects such like nucleon recoils in weak interactions. 
\end{abstract}

\pacs{14.60.Pq, 97.60.Bw}   

\maketitle

\section{Introduction} 

The detection of neutrinos from core-collapse supernovae (SNe)
represents the most exciting frontier of low-energy neutrino ($\nu$) astronomy. 
Even though galactic SNe are rare, perhaps a few per century, 
the existing  large underground neutrino detectors and the numerous planned ones
increase the confidence that a high-statistics SN neutrino signal will be eventually
observed. 
Such a detection would provide a plethora of astrophysical information
on the SN explosion mechanism, and could offer a handle
on particle physics such as $\nu$ masses and mixings, too (see, e.g.,~\cite{Raffelt:2010zz}).
  
In particular, the  flavor conversions occurring deep inside the star could leave an
imprint on the observable SN neutrino burst.
A lot of attention has been paid to  possible signatures of the
Mikheyev-Smirnov-Wolfenstein~(MSW)  effect~\cite{Mikheev:1986gs,Wolfenstein:1977ue}
with the ordinary matter in the stellar envelope~\cite{Dighe:1999bi}. 
Moreover, in recent years it has been realized that in the deepest
SN regions the neutrino density is so high that the neutrino-neutrino
 interactions~\cite{Pantaleone:1992eq,Qian:1994wh}   dominate
the flavor evolution in a highly non-trivial way (for a review see~\cite{Duan:2010bg}). 
The general result of these studies is that rapid conversions between different flavors are possible and can occur collectively, i.e. in a 
coherent fashion for many modes over large energy ranges. 
Unfortunately, the occurrence of these effects is strongly dependent on the original SN emission 
features, which makes a {\it general} characterization of the observable SN neutrino spectra 
at Earth in terms of the original ones a formidable task 
(see, e.g,~\cite{Duan:2006an,Fogli:2007bk,Dasgupta:2009mg,Duan:2010bf,Mirizzi:2010uz,Mirizzi:2011tu}). At the moment we are still far 
from a complete understanding of this complex flavor-dynamics.

However, {\it the lack of a complete understanding} should not be confused with {\it a complete
lack of understanding}. In fact, for some conditions/regimes a relatively robust comprehension has been
achieved. This is {\it not} the case, unfortunately, for the long-time cooling 
phase (postbounce time  $t_{\rm p.b.}\gtrsim 1$~s).
In principle, rich time- and energy-dependent collective dynamics may be present
 there~\cite{Duan:2010bf,Mirizzi:2010uz,Mirizzi:2011tu,Choubey:2010up}, on the top of which
peculiar time-dependent modification of the flavor content of the flux could be induced by  MSW effects
associated to the shock-wave propagation in the stellar 
envelope~\cite{Schirato:2002tg,Fogli:2003dw,Fogli:2004ff,Fogli:2006xy,Tomas:2004gr}.
Since our current understanding suggests that the resulting neutrino spectra depend on many poorly understood
details,  sharp predictions for the flavor evolution are very challenging if not impossible at present.
Furthermore, during the cooling phase all neutrino flavors originate
close to the neutron star surface, where the material is very neutron rich, suppressing charged-current reactions for ${\overline\nu}_e$. Therefore,  one
expects that the luminosities and spectra of ${\overline\nu}_e$ and $\overline{{\nu}}_x$ become quite similar, making it much harder to see flavor oscillation effects at all in the dominant ${\overline\nu}_e$  channel, which is  currently the optimal one. This is due to the fact that all large existing and near-future detectors primarily see inverse beta decay events ${\overline \nu}_e +p\to n  + e^+$.  
The relative similarity of  ${\overline\nu}_e$ and $\overline{{\nu}}_x$ spectra seems to be in fact qualitatively confirmed by recent 1D long-time simulations~\cite{Fischer:2009af,Huedepohl:2009wh,fischer}.

In order to direct future searches and experimental perspectives, a more robust strategy is to  focus on the early phase of the SN neutrino
signal, in a time window where relatively robust expectations exist for  the neutrino emission spectra and for the flavor dynamics.
The largest difference among the flavor fluxes arises during the first 10--20 ms after bounce when
the outer layers of the collapsed core deleptonize, leading to the prompt $\nu_e$ burst. Since negligible $\bar\nu_e$ and ${\nu}_x$ fluxes are emitted
during this phase, self-induced oscillations are simply absent:
collective effects do not give rise to any flavor transformation during the
neutronization burst~\cite{Hannestad:2006nj}~\footnote{An exception is constituted by the case of low-mass SNe with
an oxygen-neon-magnesium core, where the matter density profile can be so steep that the usual MSW matter 
effects occur within the region of high neutrino densities close to the neutrino sphere, triggering
self-induced flavor conversions there~\cite{Duan:2007sh,Dasgupta:2008cd}. We neglect such case in the following,
also because one expects to be able to identify this type of SNe via the characteristics (in particular the duration) of their accretion phase.}.
Additionally, at these early times, since the shock wave stalls
at low radii close to the neutrinosphere,  the MSW flavor transitions
occurring at larger radii would essentially probe the static SN progenitor profile. 
In this situation, the characterization of the flavor conversions is straightforward and,
since the model predictions for the energy and luminosity of the burst are
fairly robust, the observation of the burst gives direct information about the 
survival probability of $\nu_e$ and then on the mixing parameters~\cite{Kachelriess:2004ds}. Indeed, a strong
suppression of the $\nu_e$ burst would be a smoking gun for the normal
neutrino mass hierarchy  (NH: $\Delta m^2_{\rm atm} = m_3^2 - m_{1,2}^2>0$)
in the case of a ``large'' $1-3$ leptonic mixing angle (i.e. $\sin^2 \theta_{13}\gtrsim 10^{-3}$), as currently 
measured by the
Daya Bay~\cite{An:2012eh} and Reno~\cite{Ahn:2012nd}   reactor experiments. These
recent measurement confirm and greatly strengthen the
significance of early hints suggested by the long-baseline $\nu_\mu\to\nu_e$ experiments~\cite{Abe:2011sj,Adamson:2011qu}
and Double Chooz reactor experiment~\cite{Abe:2011fz}, especially when analysed in combination with other oscillation data~\cite{Fogli:2011qn,Schwetz:2011zk}.
However, with current ``${\overline\nu}_e$ SN detectors'', such effects are challenging-to-impossible to
detect, and one has to invoke future Mton-class water Cherenkov detectors~\cite{Kachelriess:2004ds} or large liquid-argon time projection chambers~\cite{GilBotella:2003sz},
to achieve enough sensitivity to these $\nu_e$ signal features.

On the other hand,  the subsequent phase of mass accretion, characterized by a typical postbounce timescale $t_{\rm p.b.}$ of ${\cal O}$(100) ms, 
represents a particularly interesting possibility for detecting signatures of  flavor 
transformations also in the accessible ${\overline\nu}_e$ channel
(see~\cite{Pagliaroli:2008ur} for a discussion in the context of SN1987A). First of all, one can more easily afford to simulate these early stages with
sufficiently realistic neutrino transport than the longer timescales of later cooling phases. Also, the neutrino signal properties are largely independent of the
detailed mechanism of the explosion (and actually of the question whether an explosion takes place at all), since the revival of the shock wave has yet to
take place. All modern simulations available indicate that the neutrino fluxes as well as the flavor-dependent flux differences are large in this phase, with a robust
hierarchy for the neutrino number fluxes, $F_{\nu_e} > F_{{\overline\nu}_e} \gg F_{\nu_x}$,
where $\nu_x$ indicates the non-electron flavors.
Moreover, it has been recently realized~\cite{Chakraborty:2011nf,Chakraborty:2011gd}
 that the net electron densities $n_e$  reached above the neutrinosphere in realistic SN models 
 exceed the neutrino density $n_{\nu}$,  significantly suppressing the development of the self-induced neutrino oscillations
 according to the ``multi-angle matter suppression'' mechanism first described in~\cite{EstebanPretel:2008ni}.  
The matter suppression  ranges  from complete (when $n_e \gg n_\nu$) to partial
(when $n_e \gtrsim n_\nu$), producing in principle intriguing time-dependent features.
Using as benchmark the results of the hydrodynamical SN simulations of the Basel/Darmstadt  group for different iron-core SN models~\cite{Fischer:2009af}, Ref.~\cite{Chakraborty:2011gd} found complete matter suppression for postbounce times $t_{\rm p.b.} \lesssim 0.2$~s, and partial flavor conversions for $0.2 \lesssim t_{\rm p.b.} \lesssim 0.4$~s.
This result has been independently confirmed in Ref.~\cite{Sarikas:2011jc}.
More recently, in a  work based on a iron-core SN model from the Garching group  complete matter suppression has been found for all the duration of the  accretion phase~\cite{Sarikas:2011am}. 
When the matter suppression is complete,
the $\nu$ signal will be processed only by the usual
MSW effect in the SN mantle with the static progenitor profile.
In this situation, the  characterization of the SN neutrino signal results
is straightforward~\cite{Dighe:1999bi}. Moreover, in the presence of a ``large'' $\theta_{13}$ mixing angle
 one expects significant differences in the observable
${\overline\nu}_e$ flux for the  two mass hierarchies~\cite{Dighe:1999bi}. The diagnostic power of this observable 
has been given relatively little attention so far:  the hierarchy discrimination power in IceCube was briefly addressed as side
goal in~\cite{abbasi:2011ss}, which appeared when our work was in progress, but with inadequate simulations and no discussion
of the model-dependence, as we shall comment later. On the other hand the relatively model-independent flavor difference
in rise time behaviour was already visible (limited to the first 20 ms) in the early study~\cite{Kachelriess:2004ds}, but its potential diagnostic power
was not explored there.

Motivated by these considerations, we devote our work to a characterization of the early SN neutrino lightcurve signal
in the largest current neutrino detector for such a purpose, namely the IceCube Cherenkov detector in the ice at the South Pole. We shall focus in particular
on observables sensitive to the neutrino mass hierarchy. 
The plan of our work is as follows. In Sec.~\ref{siminput} we present the input neutrino flux models we considered, obtained from recent radiation-hydrodynamical simulations from the Garching~\cite{Garchingmodels} group.
In Sec.~\ref{flavour} we characterize the neutrino flavor
conversions during the accretion phase. 
In Section~\ref{icecube} we describe the SN neutrino signal in IceCube. 
We show how  the analysis of the neutrino lightcurve during the accretion phase may be
 a powerful tool to probe the neutrino emission features and the mass hierarchy in the
likely case the mixing angle $\theta_{13}$ is not too small. We substantiate and quantify our idea with an extensive
scan via Monte Carlo simulations.  We also discuss the importance of cross-checking these results to quantify better
the theoretical error, illustrating the qualitative agreement but also delicate points to be addressed via a brief account of
analogies and differences found in three simulations by the Basel/Darmstadt group~\cite{Fischer:2009af}. Perspectives on the field and conclusions follow, in Sec.~\ref{conclusions}.

\section{Numerical models for supernova neutrino emission}
\label{siminput}
We summarize here the main aspects of the SN models considered for the present investigation. Quantitative statistical analysis are based on simulations that were performed by the SN group of Garching~\cite{Garchingmodels}, while some qualitative considerations in Sec.~\ref{discussion} will also make use of simulations from the Basel/Darmstadt group~\cite{Fischer:2009af}. In Sec.~\ref{nuemprop} a qualitative description of the features found in simulations is given. Readers  not interested in details may skip directly to Sec.~\ref{nuradprop}, where the basic input and parametrizations used in the following analysis  are summarized.

\subsection{Qualitative discussion of supernova neutrino spectra}\label{qualdisc}

The Garching group models we consider~\cite{Garchingmodels} were computed with the PROMETHEUS-VERTEX code~\cite{Ramp:2002,Buras:2006}. It contains hydrodynamics modules for both spherically symmetric (1D) and multi-dimensional simulations (using a polar coordinate grid). These are based on a conservative and explicit Eulerian implementation of a Godunov-type scheme with higher-order spatial and temporal accuracy. Although the solver is Newtonian, it employs a correction to the gravitational potential approximating effects of general relativistic gravity (case A of Ref.~\cite{Marek:2006}). The module for the energy-dependent, three-flavor neutrino transport solves the ${\cal O}(v/c)$ moment equations for neutrino energy, momentum, and lepton number with a variable Eddington-factor closure obtained from a model-Boltzmann equation. General relativistic redshifting is included and a ``ray-by-ray plus'' approximation is employed for treating multi-dimensional problems. The set of neutrino processes used for the simulations analyzed in the present work was discussed in Ref.~\cite{Buras:2006} (see Appendix~A there). It is supplemented by the improved electron-capture rates on heavy nuclei of Ref.~\cite{Langanke:2003ii} and the inelastic neutrino-nucleus scattering rates of Ref.~\cite{Langanke:2007ua}, both of which have some influence on the details of the core-infall phase before bounce and on the exact formation point of the SN shock.

The set of spherically symmetric core-collapse simulations from the Garching group is based on a selection of progenitor models from Ref.~\cite{Woosley:2002zz} in addition to the older 15\,$M_\odot$ model from Ref.~\cite{Woosley:1995ip}, for which we  also evaluate results of an axially-symmetric (2D) simulation published in Ref.~\cite{Marek:2009}.  None of the runs produced an explosion within the evolution periods considered for the present work.
All calculations were performed with the equation of state (EOS) of Lattimer \& Swesty~\cite{Lattimer:1991}, with a bulk incompressibility modulus of 180~MeV and symmetry energy of 29.3~MeV for nuclear matter. This would eventually imply a cold neutron star maximum mass of nearly 1.85 M$_\odot$. At face value, this is at odds with  the recent mass determination of $1.97\pm0.4$~M$_\odot$ from the millisecond pulsar J1614-2230 observation~\cite{Demorest:2010bx}. However, one should bear in mind that constraints on the EOS derived from neutron star masses and radii apply to the {\it cold} EOS, which are not ``per se'' representative of  the conditions of {\it hot}, more isospin-symmetric SN matter, whose description may be thus accurate with the currently used EOS. Also, in order to fulfill the above-mentioned constraint, the Lattimer \& Swesty EOS with incompressibility modulus of 220~MeV can be applied, which gives a cold neutron star maximum mass of 2.05~M$_\odot$. In SN simulations, both EOS---based on 180 and 220~MeV---result in essentially indistinguishable evolutionary conditions during the first $\sim$100~ms after core bounce, which is the time window we are interested in. Differences obtained applying the EOS with incompressibilities of 180 and 220~MeV might become more prominent only during the later post-bounce evolution if the neutron star mass is close to the limiting mass of the EOS (see the useful study~\cite{LSM:1994}). A more extended and up-to-date discussion of this point can be found in Ref.~\cite{Mueller:2012is}.

\subsection{Neutrino emission properties}\label{nuemprop}
Here we concentrate on the early postbounce evolution, where according to the current understanding non-radial hydrodynamic flows, if present, do not have a very strong influence on the properties of the neutrino radiation leaving the SN core. 
\begin{figure}[!t]  
\includegraphics[angle=0,width=1.\columnwidth]{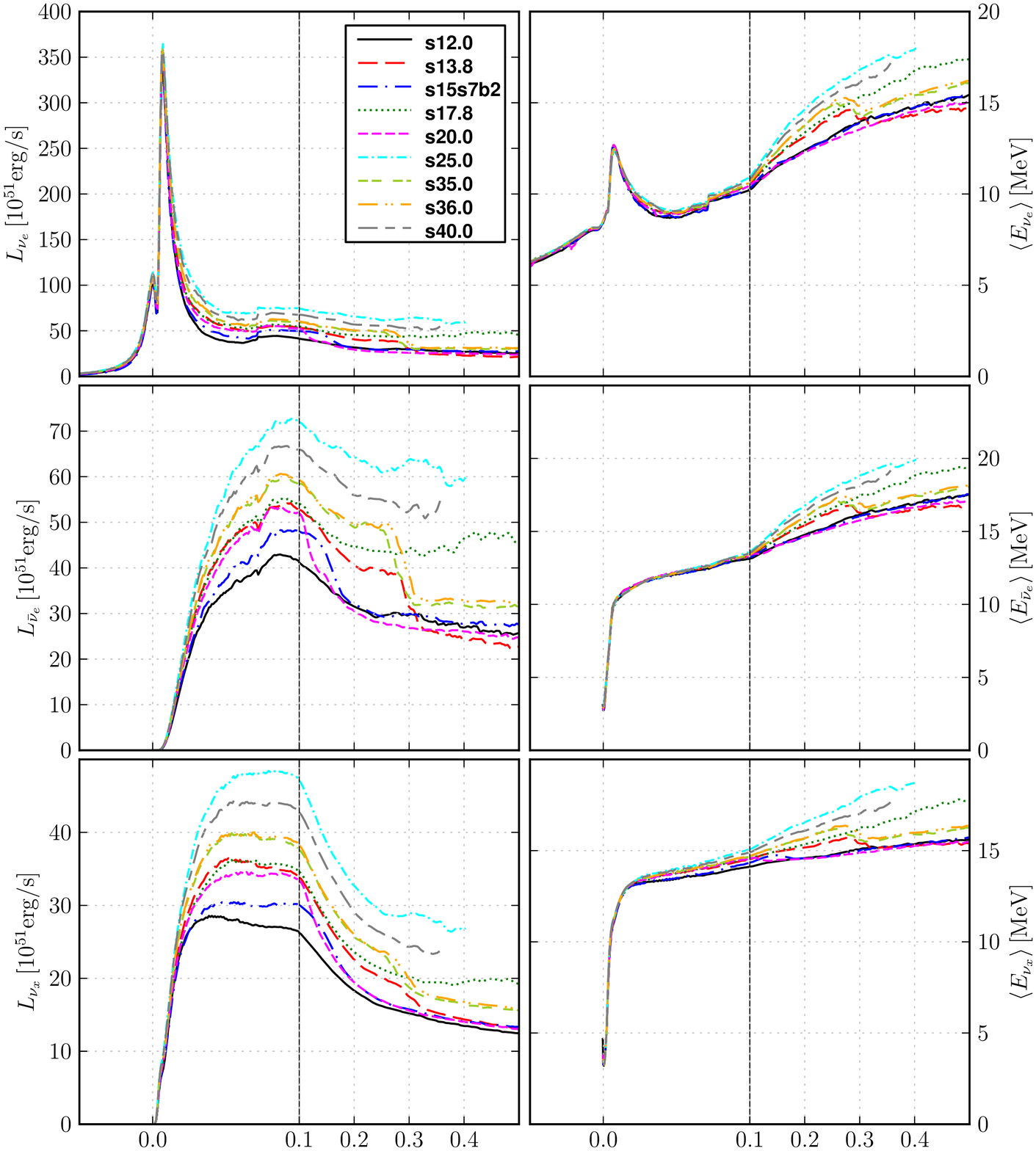} 
\caption{
Early postbounce evolution of luminosities (left panels), and mean energies (right panels)
for a set of nine 1D simulation with progenitors of different masses (see text for details) as obtained by the Garching group~\cite{Garchingmodels}.
Quantities for  $\nu_e$,   $\overline\nu_e$, and  $\nu_x$   are shown in the top, middle and bottom panel, respectively. The vertical line
indicates the early timescale (100 ms) of particular interest in this article.
\label{fig1}}  
\end{figure}  

Figure~\ref{fig1} illustrates the evolution of the neutrino luminosities and average energies $\langle E_\nu\rangle$ for $\nu_e$,
 $\overline\nu_e$ and $\overline\nu_x$ for a nine 1D models, sampled in the observer frame at infinity (for a definition of these observables, see~\cite{Liebendoerfer:2005}). Here, $\nu_x$ indicates both $(\mu,\tau)$-(anti)neutrinos. The neutrino emission properties observed at infinity are determined at the moment of decoupling, which defines the neutrino-energy and flavor dependent spheres of last scattering~\cite{Keil:2002in}. The neutrinos which contribute to the luminosity at infinity carry information about the state of the matter at the moment of decoupling.  In the following, we will discuss general properties of the observables shown in Fig.~\ref{fig1}.

During the early postbounce accretion phase all investigated SN models have a common feature, which they also share with results published in Refs.~\cite{Fischer:2009af,Liebendoerfer:2004,Liebendoerfer:2005,Marek:2009,Mueller:2010,Lentz:2011aa}: The luminosity of heavy-lepton neutrinos and antineutrinos (henceforth collectively denoted by $\nu_x$) rises to its maximum level  faster than that of electron  antineutrinos $\overline\nu_e$. Note that $\overline\nu_e$ and $\nu_x$ in contrast to $\nu_e$ are not emitted in any significant amounts during the core-collapse phase until core bounce. Instead, their vivid production sets in only when the bounce shock starts to heat swept-up material to high temperatures. This allows nucleon-nucleon bremsstrahlung to become efficient and positrons to appear so that electron-positron annihilation can also take place. These processes become more and more important as the temperature rises and the electron degeneracy drops as a consequence of the deleptonization triggered by the prompt $\nu_e$ burst. The production of $\overline\nu_e$ is more strongly suppressed than that of $\nu_x$ during the first $\sim$20\,ms after bounce because of the high degeneracy of electrons and $\nu_e$, which are present in very large numbers before and during the emission of the deleptonization burst\footnote{Since the high electron degeneracy allows only for a low abundance of positrons, the production of $\overline\nu_e$ by $e^+e^-$ annihilation and $e^+$ captures on neutrons is not efficient. Moreover, since in the optically thick regime $\nu_e$ are in chemical equilibrium with the matter their degeneracy also blocks the phase space for the creation of $\overline\nu_e$ via nucleon-nucleon bremsstrahlung.}. This was already visible in~\cite{Kachelriess:2004ds}.

The steep initial increase of the $\nu_x$ luminosity is followed by a relatively abrupt termination of this growth at a value of typically a few $10^{52}$~erg/s (for a single kind of heavy-lepton neutrino), considerably (almost a factor of two) below the peak luminosity reached by $\overline\nu_e$ more gradually about 0.1~s later. During this phase, the emission of $\nu_e$ and especially $\overline\nu_e$ grows thanks to their highly efficient production via charged-current processes (electron and positron captures on free nucleons) in the matter that forms a thick, hot mantle around the newly born proto-neutron star after having been accreted through the standing bounce shock. The transition from a growing/plateau phase to a decreasing luminosity  depends on the core structure of the collapsing star and the corresponding shallow decline of the mass-infall rate with time and thus varies with the progenitor: in the models considered it varies from about 0.1 to 0.3 s. A faster luminosity drop sets in when the density and thus mass accretion rate decreases more abruptly. This can be associated with, e.g., the infall of an interface between progenitor shells containing different chemical compositions or with the onset of the explosion, which quenches further accretion. Small oscillatory behaviors superimposed on the main trend in luminosities visible in Fig.~\ref{fig1} reflects the back- and forward propagating bounce shock, which in turn leads to an oscillating mass-accretion rate behind the SN shock.

While the production of $\nu_e$ and $\overline\nu_e$ in the hot accretion layer is very important, heavy-lepton neutrinos are created exclusively by neutral-current pair processes, i.e., nucleon-nucleon bremsstrahlung as well as electron-positron and $\nu_e$-$\overline\nu_e$ pair annihilation. Consequently, heavy-lepton neutrinos are thermally less strongly coupled to the stellar medium. Their main production occurs deeper inside the newly born proto-neutron star and additional contributions from the accretion layer are less significant. Therefore $\nu_x$ escape effectively from a layer with smaller radiating surface and their luminosity remains lower than that of electron-flavor neutrinos (since the emission is approximately blackbody-like, the luminosity $L_\nu$ scales with the neutrinosphere radius $R_\nu$ and temperature $T_\nu$ roughly like $L_\nu\propto R_\nu^2T_\nu^4$). The temporal evolution of the $\nu_x$ luminosity after the peak depends on two factors, which in combination have a complex influence on the evolution of temperature and radius of the corresponding neutrinosphere: (a) the contraction behavior of the core of the nascent proto-neutron star and (b) the growing thickness of the surrounding accretion layer in response to the continuous infall of matter from the collapsing progenitor star. On the one hand a more massive accretion layer compresses the proto-neutron star core but at the same time leads to higher temperatures, therefore not only enhancing the $\nu_e$ and $\overline\nu_e$ emission but also the $\nu_x$ luminosity of more massive progenitors. On the other hand the core contracts faster also for a softer nuclear equation of state. Therefore, both the different stellar progenitors (with a different structure of their iron core and surrounding regions) and the different equations of state have some consequences for the properties and evolution of the emission of neutrinos and antineutrinos of all flavors during the postbounce accretion phase.

It is important to note that outside the radius where pair-production processes have essentially ceased (at the so-called ``energy-sphere''), $\nu_x$ still diffuse through an overlying layer of opaque matter, in which they scatter frequently off neutrons and protons before they can escape freely at their ``transport-sphere''. Although the energy transfers from  high-energy $\nu_x$ to the nucleons of the cooler environment by individual scatterings are small, the cumulative effect of many scattering reactions adds up to a noticeable down-grading in energy space of $\nu_x$ that finally stream off the transport-sphere (for a detailed discussion, see Ref.~\cite{Raffelt:2001}). The corresponding spectral changes are accounted for, since the effects of nucleon recoil and thermal motions (as well as weak-magnetism corrections) are included in the treatment of neutrino-nucleon interactions (see Appendix~A of Ref.~\cite{Buras:2006}). 

Differently from the spherically symmetric (1D) cases shown in Fig.~\ref{fig1}, the observable neutrino luminosities and mean energies predicted by multi-dimensional SN simulations exhibit short-time variations with considerable amplitudes during the postbounce accretion phase. This is connected to the presence of large-scale asymmetries as a consequence of hydrodynamic instabilities in the layer between proto-neutron star surface and stalled bounce shock. These lead to time-dependent hot spots and anisotropic neutrino production in the cooling region of the settling accretion flow. Such effects were discussed in detail for axially symmetric (2D) models in Refs.~\cite{Marek:2009,Ott:2008} and for 3D models in Ref.~\cite{Muller:2011yi}. Apart for these subleading
features, there is no significant qualitative difference in multidimensional simulations.

\subsection{Parameterization of neutrino radiation properties}\label{nuradprop}
In summary, numerical simulations of core-collapse SNe provide the un-oscillated doubly differential neutrino distribution in energy and time,
\begin{equation}
F^0_\nu\equiv\frac{d^2 N_\nu}{d t \,d E}\,.\label{eq:nudistrib}
\end{equation}
where $\nu=\left\{ \nu_e, \overline\nu_e, \overline\nu_x\right\}$ in standard notation~\cite{Dighe:1999bi}. This is related to the \emph{instantaneous} (time-dependent) luminosity via
\begin{equation}
L_\nu=\int_{0}^{\infty} dE E F^0_\nu  \,\ .
\end{equation}

We factorize simulation outputs as follows:
\begin{equation}
F^0_\nu = \frac{dN_\nu}{dt}\varphi(E_\nu)
\end{equation}
for each flavor ($\nu=\nu_e, \overline\nu_e, \nu_x$), where 
\begin{equation}
\frac{dN_\nu}{dt} = \frac{L_\nu}{\langle E_\nu\rangle}
\end{equation}
represents the neutrino emission rate (number of $\nu$'s per unit of time) with mean neutrino energy $\langle E_\nu\rangle$. The function $\varphi(E)$ is the normalized  ($\int \varphi(E)dE = 1$) energy spectrum parametrized as in~Ref.~\cite{Keil:2002in}
\begin{equation}
\varphi(E)= \frac{1}{\langle E_\nu\rangle}\frac{(1+\alpha)^{1+\alpha}}{\Gamma(1+\alpha)}\left(\frac{E}{\langle E_\nu\rangle}\right)^\alpha\exp\left[-(1+\alpha)\frac{E}{\langle
E_\nu\rangle}\right]\, ,
 \label{eq:varphi}
\end{equation}
where the energy-shape parameter $\alpha$ is defined as~\cite{Keil:2002in,Raffelt:2001}
\begin{equation}
\alpha=\frac{2\langle E_\nu \rangle^2-\langle E_\nu^2\rangle}{\langle E_\nu^2\rangle-
\langle E_\nu\rangle^2} \,, \label{alphadef}
\end{equation}
i.e. it is a dimensionless parameter containing information on the second moment of the distribution, $\langle E_\nu^2\rangle$. 
 In general,  $L_\nu$, $\langle E_\nu\rangle$ and $\alpha$ are all functions of time, and are  extracted directly from the simulations at hand.

\begin{figure}[t]  
\begin{center}
\includegraphics[angle=0,width=0.7\columnwidth]{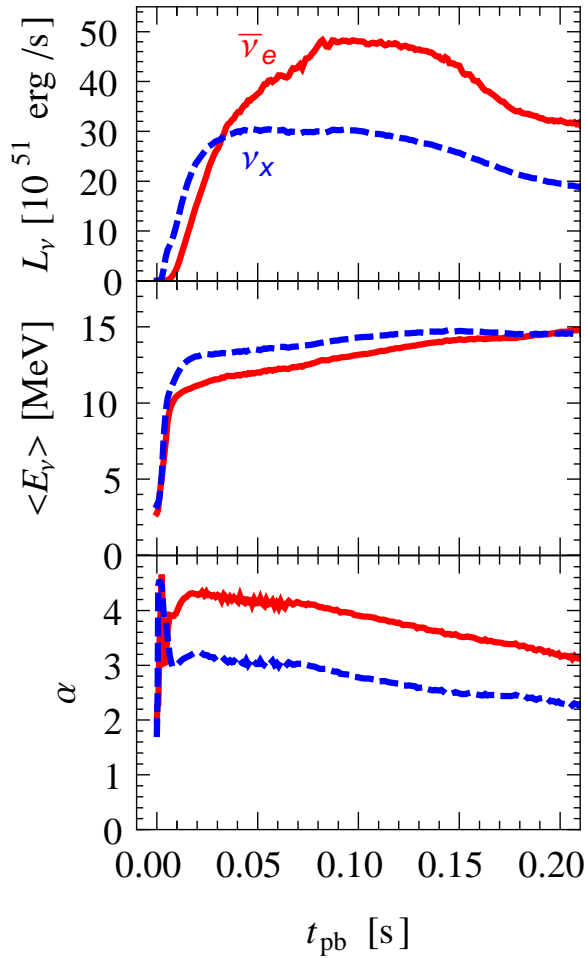} 
\caption{
Early postbounce evolution ($t_{\rm p.b.} \leq 0.2\,$s) of luminosities (top panel), mean energies (middle panel)
and $\alpha$-fit parameters (bottom panel) for the 15~$M_\odot$ progenitor (see text for details), for $\overline{\nu}_e$ (solid lines) and $\nu_x$ (dashed lines) species.
\label{fig2}}  
\end{center}
\end{figure}  
We shall be essentially interested in $\overline{\nu}_e$ and $\nu_x$ observables at early times, i.e. $t_{\rm p.b.} \leq {\cal O}(0.2)$~s (for illustrative purposes, we reproduce the key variables in this time frame for  the 15~$M_\odot$ progenitor in Fig.~\ref{fig2}.) In this time window of the pre-explosion phase the non-radial mass motions connected to hydrodynamical instabilities (i.e. hot-bubble convection and the standing accretion shock instability) found in multi-dimensional SN models have still to grow to their full strength~\cite{Muller:2011yi}. Indeed, comparing the neutrino luminosities and mean energies of the 15 $M_{\odot}$  1D model of Fig.~\ref{fig2} with the corresponding quantities for the 15 $M_{\odot}$ 2D model in Fig.~6 and 7 of Ref.~\cite{Marek:2009}, one realizes that during the early accretion phase only minor differences arise between these two cases.
Therefore, the 1D SN models we will use for our study are sufficient for an accurate characterization of the neutrino signal and of the SN matter density at early postbounce times.
We will show this explicitly in Sec.~4 where we will compare our results for 1D and 2D SN models. 
This similarity between 1D and 2D results also simplifies the neutrino flavor conversion physics, as explained in the following section.

\section{Neutrino flavor conversions}\label{flavour}

The emitted SN neutrino flux is processed by self-induced and
MSW oscillation effects during its propagation. 
The self-induced effects would take place within $r \sim \mathcal{O}(10^{3})$~km from the neutrinosphere
whereas the MSW transitions take place at larger radii, in the region $r \sim 10^{4}$--$10^{5}$~km. 
As the self-induced and MSW effects are widely separated in space, they can be considered
independently of each other.
In the normal mass hierarchy (NH, $\Delta m^2_{\rm atm}>0$) and for the spectral ordering of the accretion phase,
no self-induced flavor conversion  will  occur.
Instead, in inverted mass hierarchy (IH, $\Delta m^2_{\rm atm}<0$) potentially large self-induced effects could be expected~\cite{Mirizzi:2010uz}. 
However, it has been shown using results both  from Basel/Darmstadt group simulations and Garching group ones~\cite{Chakraborty:2011nf,Chakraborty:2011gd,Sarikas:2011am} that the trajectory-dependent  multi-angle effects associated with the dense ordinary matter suppress collective oscillations in actual models of iron-core SNe, as expected when large trajectory-dependent  phase dispersion are induced by the matter would suppress the collective phenomena~\cite{EstebanPretel:2008ni}. 

In principle, depending on the electron density, the matter suppression can be complete, when $n_e\gg n_\nu$, or partial when the matter dominance is less pronounced. Forward-peaked angular distributions of the neutrino field further reduce the effective neutrino-neutrino potential strength and make the
effect more prominent. In the following we will focus on the early time $t_{\rm p.b.}<t<{\cal O}\,$(0.2) s where the matter density largely exceeds the neutrino density, and hence also completely suppresses the collective oscillations for the cases under investigation, as explicitly checked and confirmed by two independent groups~\cite{Chakraborty:2011nf,Chakraborty:2011gd,Sarikas:2011am}.
  
In this situation, the neutrino fluxes can only undergo the  traditional MSW conversions
in SN while passing through the outer layers of the star.
Therefore,  it is straightforward how to calculate the
$\overline\nu_e$ flux at Earth in the different cases~\cite{Dighe:1999bi}. 
In particular, in NH one finds
\begin{equation}
F_{\bar\nu_e} = \cos^2\theta_{12} F^0_{\bar\nu_e} + \sin^2\theta_{12}
 F^0_{\bar\nu_x} \,\ ,
\label{eq:nh} 
\end{equation}
where $\theta_{12}$ is the 1--2 mixing angle, with 
$\sin^2 \theta_{12}  \simeq 0.31$~\cite{Fogli:2011qn}.
In IH for ``large'' $\theta_{13}$ (i.e. with $\sin^2 \theta_{13}
\gtrsim 10^{-3}$)  one gets
\begin{equation}
F_{\bar\nu_e} = F^0_{\bar\nu_x} \,\ ,
\label{eq:ih}
\end{equation}
while for ``small'' $\theta_{13}$ (i.e. with $\sin^2 \theta_{13}
\lesssim 10^{-5}$) one finds
\begin{equation}
F_{\bar\nu_e} = \cos^2\theta_{12} F^0_{\bar\nu_e} + \sin^2\theta_{12}
 F^0_{\bar\nu_x} \,\ .
\label{eq:ihsmall}
\end{equation}
Then, it is clear that for ``large'' $\theta_{13}$ the ${\overline\nu}_e$ flux at the Earth is basically reflecting the original $F_{\bar\nu_x}$ flux, if IH is realized, or closely matching the $F_{\bar\nu_e}$ flux, in case of NH. Since these two extremes show significant qualitative differences (see Fig.~\ref{fig1}), one might hope to be able to  distinguish these two possibilities. We shall see in the next section that this appears to be a promising perspective for neutrino detection  at IceCube.

Note that in principle the signal may be further modified by Earth matter effects, whenever the SN is seen through a significant chord crossing the Earth, below the horizon. For the South Pole location, the probability for such a crossing has been estimated to be about 40\% (but just 6.4\% for a Earth core crossing) see~\cite{Mirizzi:2006xx}. In case of Earth ``shadowing'', the expected variation of signal in the accretion phase was already estimated in~\cite{Dighe:2003be} to few percent at most. We confirmed with explicit calculations in several instances that the variations of the signal are comparable to this level or smaller, and that to a significant extent they affect the overall normalization and thus are challenging to detect. In any case, these modifications are irrelevant for the type and size of the effects we focus on and will be neglected in the following.

\section{Neutrino lightcurve in IceCube}~\label{icecube}
The idea of using giant high energy $\nu$ under-water or under-ice detectors as SN neutrino observatories has been proposed a long time ago, see~\cite{Pryor:1988,Halzen:1994xe}. The method is based on a sudden, correlated increase in the photomultiplier count rate on a timescale on the order of 10~s (see Ref.~\cite{abbasi:2011ss} for a recent description).
  
In its completed configuration and with its data acquisition system, IceCube
with its 5160 optical modules~\cite{abbasi:2011ss},
 has about 3 Mton
effective detection volume, representing the largest current detectors
for SN neutrinos.
The SN neutrinos streaming through the antarctic ice interact mostly
through ${\overline \nu}_e +p\to n  + e^+$ reactions.
While
fine-grained detectors, like Super-Kamiokande, reconstruct individual neutrinos on
an event-by-event basis, IceCube only picks up the average Cherenkov glow of the ice. 
To estimate the detection rate we closely  follow Refs.~\cite{Dighe:2003be,Halzen:2009sm}. The only
change consists in replacing the product of Eq.~(1) and Eq.~(6) in~\cite{Dighe:2003be} with
the following rate of energy deposition per proton
\begin{equation}
{\cal R}_{\bar\nu_e}=\int_0^\infty d E\,F_{\bar\nu_e}\,E_{\rm rel}(E)\,\sigma(E)\,,
\end{equation}
 with $E_{\rm rel}(E)$ being the energy released by a neutrino of energy $E$ and $\sigma(E)$ the neutrino-nucleon inverse beta-decay cross section,
 which we implement following~\cite{Strumia:2003zx}; the fluxes $F_{\bar\nu_e}$ are obtained from the models considered (see Eqs.~(\ref{eq:nudistrib}-\ref{eq:varphi})) and
 account for neutrino oscillations via Eqs. (\ref{eq:nh}), (\ref{eq:ih}) for the NH and IH cases, respectively. All other detector parameters (average quantum efficiency, effective photo cathode detection area, angular acceptance range, number of useful Cherenkov photons per deposited neutrino energy,  average lifetime of Cherenkov  photons) have been fixed to the fiducial values adopted in~\cite{Dighe:2003be}, which we address to for further details.

\begin{figure}[!t]  
\includegraphics[angle=0,width=0.49\columnwidth]{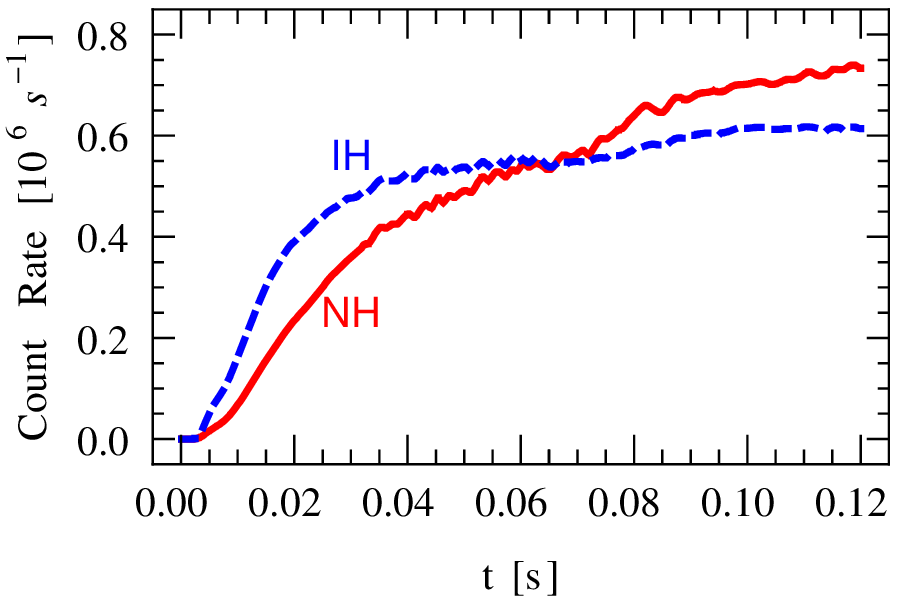} 
\includegraphics[angle=0,width=0.49\columnwidth]{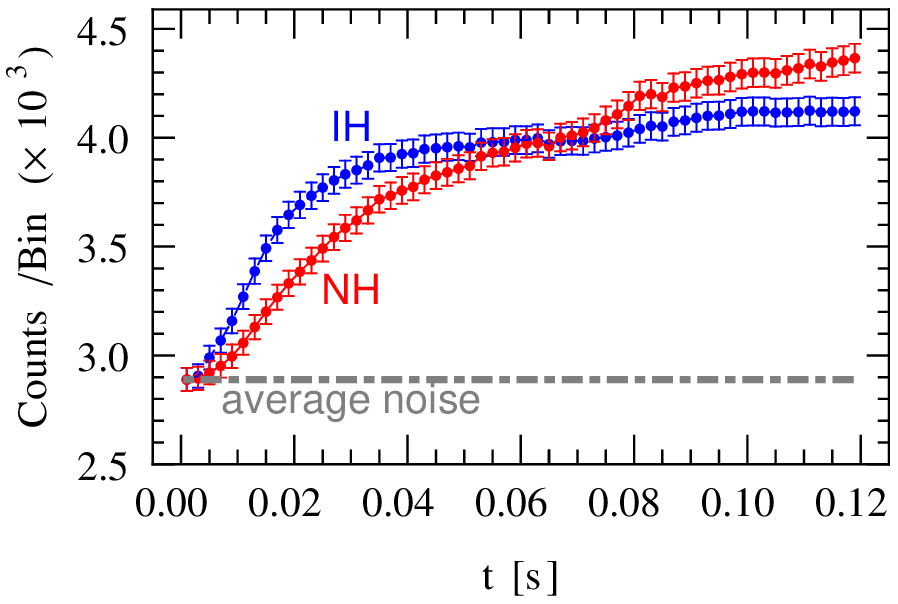} 
 \caption{Left Panel: average SN count rate signal in IceCube assuming a distance of 10 kpc,  
based on the simulations for a  15 $M_{\odot}$ progenitor mass from the Garching group.
Right panel: illustrative example of the binned signal using 2~ms bins with typical Poisson error estimates accounting for the signal plus photomultiplier background noise, whose average value is shown as dot-dashed curve. A large $\theta_{13}$ is assumed here and in the following  (see text for details).
\label{fig3}}  
\end{figure}  

In Figure~\ref{fig3} we show the expected {\it overall} signal rate $R(t)$ in IceCube for a galactic SN at a distance of $d=10$~kpc. This ``fiducial'' distance has been chosen consistently with average distance expectations~\cite{Mirizzi:2006xx} for different SN models. Since we are interested in the possibility of the mass hierarchy discrimination, we focus on the likely case of a large $\sin^2\theta_{13}$ where the observable $\overline\nu_e$ signal would be different in the two hierarchies [see Eqs.~(\ref{eq:nh})--(\ref{eq:ih})].  We refer to the cases of  15 $M_{\odot}$ progenitor masses previously used for illustration. The NH cases are shown with continuous curves, while the IH cases are the dashed curves. The right panel overlies the error size of a binned signal using 2~ms bins with typical error estimates from the photomultiplier background noise, i.e. 280 s$^{-1}$ in each optical module~\cite{Halzen:2009sm}. A large $\theta_{13}$ is assumed here and in the following.  

The difference between the observed neutrino lightcurve in the NH vs. IH is evident. Note how the NH case continues to grow steadily over the considered timescale, while the IH signal reaches quite quickly an almost constant count rate. In the case of  IH, the lightcurve has a more sudden rise. Note that both luminosity behavior and trend of growing energy of $\overline{\nu}_e$ shown in Fig.~\ref{fig1} contribute to the final shape of the curves.  Also, note that despite the relatively large differences existing over very early timescales
(10-20 ms, as already shown in~\cite{Kachelriess:2004ds}), one can already expect that integrating the signal over a longer timescales
will be needed to beat statistical errors.

\begin{figure}[!th]
\begin{center}
\includegraphics[width=0.49\textwidth]{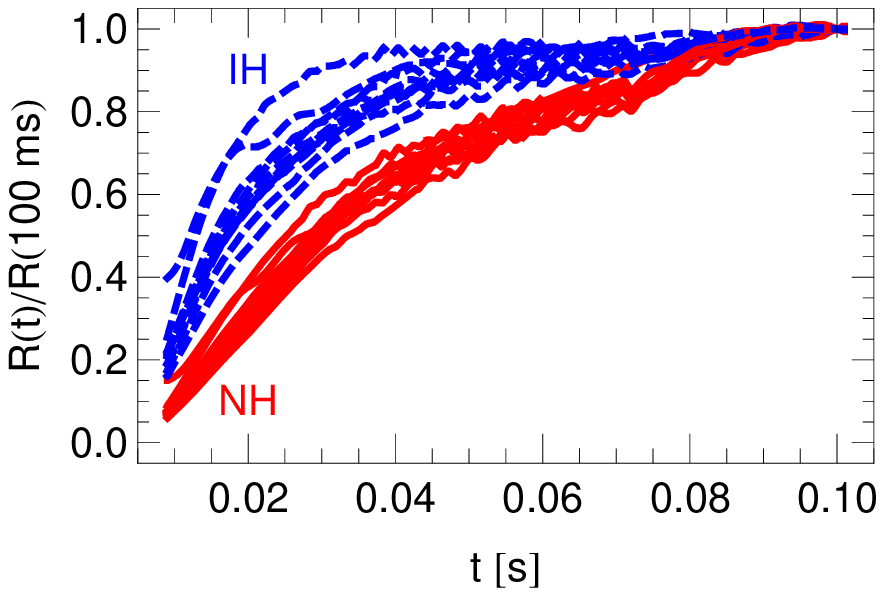}
\includegraphics[width=0.49\textwidth]{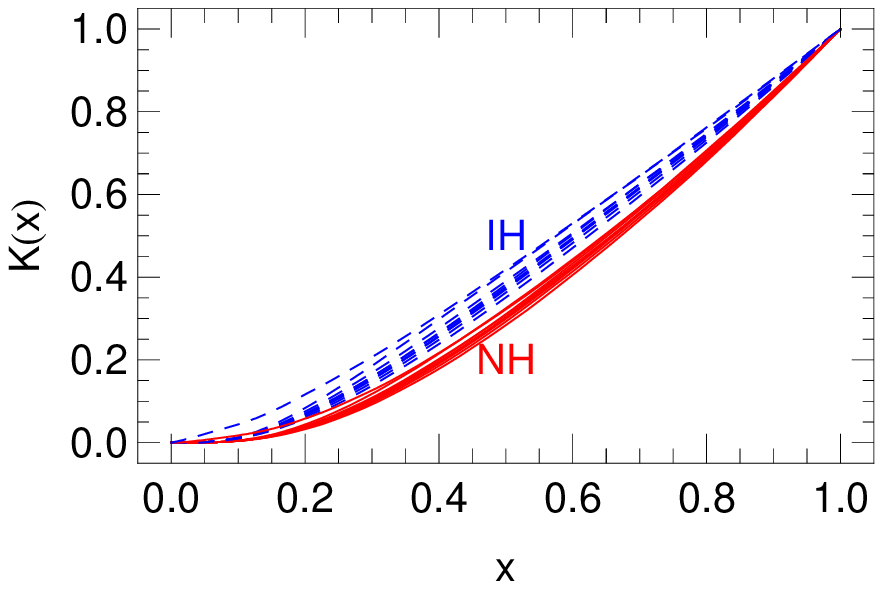}
\caption{Count rate functions $R_i(t)$ and
the cumulative functions $K_i(x)$ for the Garching set of models; dashed (top) curves denote expectations for IH, solid (bottom) curves for NH. In these examples, $t_{\rm end}=100\,$ms is assumed.}
\label{fig4}
\end{center}
\end{figure}

It is useful to compare the analogous behaviors for the whole set of models, a task which will be made easier by a(n irrelevant) rescaling to the rate measured
at the end of the time interval considered, $R(t)/R(t_{\rm end})$. Also, for the following statistical analysis, it is useful to introduce cumulative time
distributions $K(x)$, defined in terms of $R(t)$ as
\begin{equation}
K(x)=\frac{\int_0^{x\,t_{\rm end}} dt\,R(t)}{\int_0^{t_{\rm end}} dt\,R(t)}\,,
\end{equation}
which is a dimensionless function satisfying $K(0)=0$, $K(1)=1$, with $x\in [0,1]$.  In Fig.~\ref{fig4}, we illustrate the count rate functions $R_i^A(t)$ and
the cumulative functions $K_i^A(x)$ for the different models considered, with $i=1,\ldots,N\equiv 10$ labeling the simulation and $A$ (or in general capital latin letters) being the index related to the hierarchy, i.e. $A=$NH (red, bottom curves) or $A=$IH (blue, top curves).
In particular, we used the nine 1D SN models shown in Fig.~1 and a  2D SN model with a 15 $M_{\odot}$ progenitor mass. 
 Note that the difference between the two hierarchies is a {\it shape} difference (as should be clear already from Fig.~\ref{fig1}), rather than a mere overall difference in average energies, for example,  as in some past proposals for SN physical diagnostics. Also note that this difference is quite independent of the progenitor used (most notably of its mass) and, in agreement with expectations,  do not show a significant dependence from the dimensionality of the simulation either.

\subsection{Metric in Function Space}~\label{metric}

We now turn to assigning a {\it quantitative meaning} to the distance among models. To that purpose, we must introduce some metric
in the function space. We choose the so-called ${\mathcal D}_\infty$ metric, so that the distance between the predictions (always a number
between 0 and 1) writes:
\begin{equation}
{\mathcal D}_\infty(K_{i}^{A}\,, K_{j}^{B})=\max_{x\in[0;1]}\, \left|K_{i}^{A}(x)-K_{j}^{B}(x)\right |\,.
\label{eq:dist}
\end{equation}
This choice is solely dictated by the standard practice in experimental physics to use Kolmogorov--Smirnov  statistic (which uses that metric) to test whether two underlying one-dimensional distributions differ. We emphasize, however, that alternative choices are possible and in fact may lead to better discrimination power. Thus, the following results
are  to be meant as illustrative. Generically, we find  that  
``typical'' distances of a model $K_{i}^{A}$ from the models having the same hierarchy (same ``$A$'') but different simulation (different ``$i$'') is smaller than  ``typical'' distances from curves having opposite hierarchy  (no matter if with same or different $i$). To  quantify this statement, one has to  define {\it global} estimators characterizing the distance between a given curve and the whole set of distributions sharing the same or the opposite hierarchy. For example, we define
\begin{eqnarray}
\langle d\rangle(K_{i}^{A}\,, B)&=&\frac{1}{N}\sum_{j=1}^N{\mathcal D}_\infty(K_{i}^{A}\,, K_{j}^{B})\:\:\:\:(B\neq A)\,,\\
\langle d\rangle(K_{i}^{A}\,,A)&=&\frac{1}{N-1}\sum_{j\neq i}{\mathcal D}_\infty(K_{i}^{A}\,, K_{j}^{A})\,,\\
d_{\rm min}(K_{i}^{A}\,, B)&=&\min_{j} {\mathcal D}_\infty(K_{i}^{A}\,, K_{j}^{B})\:\:\:\:(B\neq A)\,,\\
d_{\rm min}(K_{i}^{A}\,,A)&=&\min_{j\neq i}{\mathcal D}_\infty(K_{i}^{A}\,, K_{j}^{A})\,,
\end{eqnarray}
namely the {\it average} distance computed from all the models or the {\it minimum} distance among all models, either of the different or
like hierarchy of the case at hand. Note that in the latter case we exclude  from the templates the model we compute the distance from. In principle, both the above functions can be used as hierarchy estimators and of course  other choices are possible. Again, we insist on the fact that here we do not
 look for an optimization of the method, we just report some examples of operational procedures to illustrate its viability. Anyway, no matter what indicator
for the distances of the curve from  the whole set of distributions is used, a natural criterion to
establish if a model $K_{i}^{A}$ is indicative of NH or IH is to evaluate wheter its distance estimator with respect to the NH class or IH class
of models is smaller. We thus define the differences 
\begin{eqnarray}
\Delta_{\rm min}(K_{i}^A)&\equiv& d_{\rm min}(K_{i}^A\,, B)-d_{\rm min}(K_{i}^A\,, A)\,,\label{Deltamin}\\
\langle\Delta\rangle(K_{i}^A)&\equiv& \langle d\rangle(K_{i}^A\,, B)-\langle d\rangle(K_{i}^A\,, A)\,,\label{Deltaav}
\end{eqnarray}
whose positive values indicate preference for the ``correct hierarchy", negative values for the ``wrong hierarchy'' with respect to the hierarchy of the 
considered model. Note that, as it will happen in practice, we assume that the experimenter has a set of templates to compare a signal with, which however {\it does not} include the correct curve. Also, ``larger'' differences obviously mean more clear separation. Of course if a difference of distances is  ``large''  or not is not an absolute concept,
but is linked to the experimental resolving power, which in turn is linked both the astrophysical factors (e.g. SN distance, energetics, etc.) and to instrumental limitations.
We shall present some quantitative estimates in Sec.~\ref{stat} and further discuss this point in Sec.~\ref{discussion}.
  
In the Tables~1 and~2 we report the results of such a computation for the choices $t_{\rm end}=100\,,120\,$ms, respectively: For each model (row index $i$: 40 M$_\odot$, etc.) we list the estimators assuming either IH (2 and 3 columns) or NH (last 2 columns).  The fact that the differences in the columns are always positive means that no matter which curve we take, it is always closer to one template of the same hierarchy type than to all templates of the opposite one, as well as always closer to the average of the  templates of the same hierarchy than to the average of templates of the opposite one. Put otherwise, within the limitations of the sample at our disposal and in the limit of no experimental errors, we can conclude that both estimator are theoretically unambiguous, or that the sets of curves are well separated.

\begin{table}[!h]
\begin{center}
\begin{tabular}{|c||c|c||c|c||}
\hline
$i$    & $\Delta_{\rm min}(K_{i}^{\rm IH})$& $ \langle\Delta\rangle(K_{i}^{\rm IH})$     & $\Delta_{\rm min}(K_{i}^{\rm NH})$ & $\langle\Delta\rangle(K_{i}^{\rm NH})$ \\
\hline
\hline
40 & 0.032 & 0.037 & 0.051 &  0.074\\
36 & 0.049 & 0.052 &  0.050 & 0.072\\
35 & 0.050 & 0.052 &  0.049 & 0.070\\ 
25 &  0.015 & 0.015 &  0.058 & 0.076\\
20 & 0.046 & 0.055 & 0.047 &   0.069\\
17.8 & 0.046 & 0.050 & 0.048 & 0.069\\
15*  &  0.062 &  0.068 &  0.027 &  0.040 \\
15 & 0.054 & 0.064 &  0.036& 0.061\\
13.8 & 0.052 & 0.060 &  0.032& 0.057\\
12 & 0.066 & 0.072 &  0.023& 0.037\\
\hline
\end{tabular}
\caption{Distance of the simulated model indicated in the first column from the set of models having
opposite hierarchy minus the one from those have the same hierarchy. Columns 2-3 assume the reference model has IH, columns 4-5
that it has  NH.  Columns 2 and 4 use the estimator defined in Eq.~(\ref{Deltamin}), columns 3 and 5 use the estimator defined in Eq.~(\ref{Deltaav}). The asterisk indicates a 2D simulation. The table assumes $t_{\rm end}=$100 ms.}
\end{center}\label{tableDist100}
\end{table}

\begin{table}[!h]
\begin{center}
\begin{tabular}{|c||c|c||c|c||}
\hline
$i$    & $\Delta_{\rm min}(K_{i}^{\rm IH})$& $ \langle\Delta\rangle(K_{i}^{\rm IH})$     & $\Delta_{\rm min}(K_{i}^{\rm NH})$ & $\langle\Delta\rangle(K_{i}^{\rm NH})$ \\
\hline
\hline
40 & 0.030 & 0.037 &  0.051 &  0.073\\
36 & 0.049 & 0.053 &   0.049 & 0.070\\
35 & 0.050 & 0.054 &  0.047 & 0.068\\ 
25 &  0.017 & 0.017 &  0.059 & 0.075\\
20 & 0.050 & 0.059 & 0.039 &   0.060\\
17.8 & 0.050 & 0.053 & 0.044 & 0.066\\
15*  &  0.062 &  0.068 & 0.027 &  0.041 \\
15 & 0.051 & 0.061 &  0.046& 0.066\\
13.8 & 0.055 & 0.062 &  0.035& 0.056\\
12 & 0.060 & 0.071 &  0.025& 0.038\\
\hline
\end{tabular}
\caption{Same as Table 1, now assuming $t_{\rm end}=$120 ms.}
\end{center}\label{tableDist120}
\end{table}

\subsection{Statistical Analysis}~\label{stat}

Next, we turn to the issue of estimating if the theoretical separations found above are ``large'' or  ``small'' compared to the expected experimental errors.
Note that, in principle, given a discrete cumulative distribution of data, the KS test may be used to estimate the probability that the measured values are drawn from any of the 
template distributions, i.e. to select (exclude) some models as compatible (incompatible) with the data.
For our purposes, however, such demand would be overly restrictive. In fact, we do not aim at identifying {\it the} correct model (i.e. the correct $A$ {\it and} $i$), but only to identify the correct class of models (i.e. the same value for $A$). To achieve this goal it is sufficient that the capability to measure the cumulative distribution is significantly better than the difference of distances obtained in the  two hierarchies, which is of the order reported in the Table~1 (Table~2), i.e. between 1.5\% and 7.6\% (between 1.7\% and 7.5\%) of the overall counts obtained within $t_{\rm end}=100\,$ms (120 ms). A back-of-the-envelope computation already shows that, at least for typical Galactic SN distances and energetics, statistical errors do not appear to spoil the viability of this method.
For fiducial parameters, we estimate  about 50000 signal photon counts in 100 ms, with almost three times higher background events. Assuming (as reasonable from the visual inspection of the plots) that most of the discrimination power comes from the middle of the interval, i.e. $x\simeq 0.5$, one can estimate a Poisson statistical uncertainty of about 0.006, or 0.6\% of the total signal count. By comparing this to the distance differences values reported in  table~1, we can tentatively conclude that the statistical errors are not expected to be the limiting factor, unless the SN is too far away, of course.

\begin{figure*}[!ht]
\begin{center}
\includegraphics[width=0.55\textwidth]{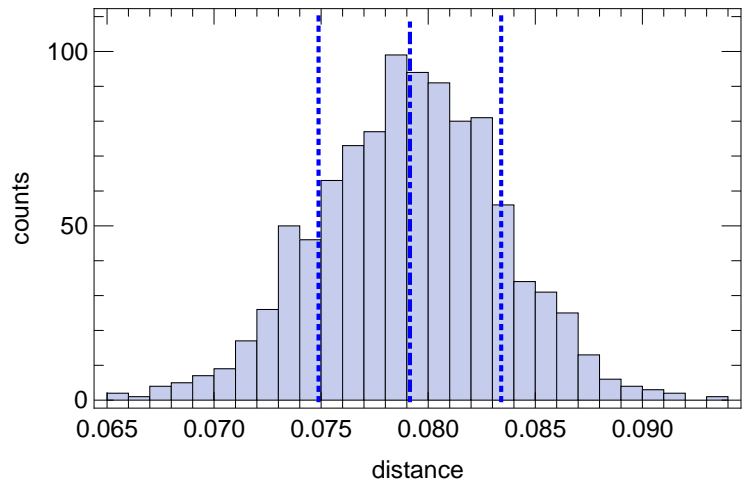}
\caption{Distribution of distances between the NH theoretical expectations for the 15
M$_\odot$ model (1D) and 1000 simulations of the IH case.
Dot-Dashed line is the average, dashed line enclose the 1-sigma range (they
enclose 68\% of the simulation results).}
\label{fig5}
\end{center}
\end{figure*}

As a check of the substantial correctness of the above argument, we performed the following exercise.
For the specific case of the 15 M$_\odot$ progenitor simulation (1D) considered
in Fig.~\ref{fig3}, the distance between the NH hierarchy model and the IH hierarchy
model is computed to be 0.078. We run 1000 Monte Carlo simulations for the IH
template, and computed the actual distances found with respect to the
``theoretical'' NH case. The histogram of the results are shown in
Fig.~\ref{fig5}, together with the average distance and the 1-sigma range
enclosed in dashed lines. Note how the statistical dispersion is in fact
comparable in size with the above estimate (actually a bit lower).

Finally, on the light of the previous encouraging results, we turn to estimating the {\it reliability} of a statistical determination
of the hierarchy along the lines previously described.  Namely, in a large number of simulations, how many times would one identify the correct hierarchy? To that purpose, we run a more extensive set of simulations, with 4000 binned histograms simulated with 2 ms bin size, accounting for noise (signal+background are assumed to be Poisson distributed in each bin). This corresponds to 400 simulations for each choice of $A$ and $i$ (hierarchy/astrophysical model). We then count the fraction of times that each of the estimators of Eq.~(\ref{Deltamin}) and
 Eq.~(\ref{Deltaav}) is smaller than zero, for the case where a NH-type curve is assumed or a IH-type of curve is assumed. These fractions, in percent, are denoted with a $\delta$ and reported in Table~3, for the two estimators introduced above and the two choices $t_{\rm end}=100\,,120\,$ms.

\begin{table}[!h]
\begin{center}
\begin{tabular}{|c||c|c||c|c||}
\hline
$t_{\rm end}$ [ms]  &   $\delta_{\rm min}({\rm NH})$ &  $\delta_{\rm min}({\rm IH})$& $ \langle\delta\rangle({\rm NH})$     &  $\langle\delta\rangle({\rm IH})$ \\
\hline
\hline
100  &  0\% & 0\% &  0\% &  0.9\%  \\
120 & 0\% & 0\% &  0.025\% & 0.5\% \\
\hline
\end{tabular}\label{tableDelta}
\caption{Reliability of the hierarchy identification. The fraction $\delta$ of wrongly identified hierarchy for the various cases is indicated.}
\end{center}
\end{table}

These results clearly show that $\Delta_{\rm min}$ is a better estimator than $\langle\Delta\rangle$: while in all cases
one can estimate a reliability of hierarchy attribution better than 99\%, by using $\langle\Delta\rangle$ the reliability improves  by at least one order of magnitude~\footnote{Due to limited statistics, we cannot exclude $\delta_{\rm min}\lesssim 0.1\%$, of course. Here we are not interested to more accurate estimates, given the illustrative purpose of our analysis with no goal of optimization.}.
Actually, the few cases where there is disagreement between the two methods are those where, due to $2-3\,\sigma$ statistical fluctuations, the separation is not too large compared with the statistical error. Getting a consistent result with different estimators may thus give a feeling for the statistical robustness of an eventual detection.

Additionally, the example above (and others not shown) also prove that the exact choice of $t_{\rm end}$ is not crucial. As a general rule, however, one should  avoid using a too small or a too large $t_{\rm end}$: despite significant separation between expectations, too small values would imply insufficient statistics and do not help; by choosing too large values of $t_{\rm end}$ would make less and less reliable the theoretical separation, both because the matter multi-angle suppression effect holds better and better at early times, and because astrophysical details (e.g. neutrino productions from outer layers of the collapsing stars) become comparatively more important at later times, especially for the NH case which is more sensitive to the accreting matter properties. 

We also note that we implicitly assumed that the ``initial signal time'' is known with negligible error with respect to the rise time shape needed to compute the function space distances. The typical computed error  with which one expects the post-bounce timing of a benchmark SN event to be established in IceCube is of about $\pm 3.5\,$ms at 95\% CL~\cite{Halzen:2009sm}. We estimated that this should translate into an uncertainty in the distances comparable with the width induced by statistical fluctuations (see e.g. Fig.~\ref{fig5}) and thus smaller than the model-to-model variation. Since our analysis suggests that this is not expected to be the limiting factor in the discrimination, a fortiori
we can guess that this should remain true when accounting for the error on the post-bounce time determination. This is also reasonable from a graphical inspection of the right panel in Fig.~\ref{fig3}, where the bin size is of 2 ms.

A future improvement over the present analysis would be to perform a simultaneous determination of the post-bounce time and the rise time shape, combining the output of analyses like~\cite{Halzen:2009sm} and the present one. This kind of development will be needed for assessing for example what is the maximal distance out to which 
a sufficiently clear separation between the two hierarchies can be achieved, with a more complete account of the detector performance. In the same spirit, another improvement may be to include elastic scattering events on electrons, which here (as in most of the literature, see for example as~\cite{Halzen:2009sm})  have been neglected since its cross-section is sub-leading by about two orders of magnitude with respect to the inverse-beta decay one.

The hierarchy discrimination power using a much more sophisticated simulation of the detector was recently estimated as a side result in~\cite{abbasi:2011ss}. 
No quantitative comparison with our results is however possible, since the simulations used in~\cite{abbasi:2011ss} only included one outdated Fe-core collapse model (from Livermore group~\cite{Totani:1997vj}) and two O-Ne-Mg progenitors from the Garching group. As we commented in the introduction, O-Ne-Mg progenitors are poorly suitable for hierarchy determination (having a short accretion phase, smaller differences between flavors, and suffering from a difficult assessment of collective effects); this expectation is in broad agreement with results of~\cite{abbasi:2011ss}, obtained however under over-simplified
assumptions for the flavor evolution. On the other hand the Livermore model, apart for showing qualitatively  that Fe-core progenitors are more promising than O-Ne-Mg  ones and that they might allow a significant discriminations power up to distance of 10 kpc or so (consistently with what illustrated here) cannot be trusted at a quantitative level due to the lack of microphysics of sufficiently sophistication, compared with present state-of-the art. Note also that the ``right model signal shape'' was assumed to be known in~\cite{abbasi:2011ss}, while in this article we show that discrimination should be possible also relaxing this assumption in a statistical sense. 

\subsection{Discussion}\label{discussion}
An important issue is how dependent this signature is from theoretical uncertainties, most notably the progenitor structure, equation of state, and numerical schemes. The set of models considered here provide quite a satisfactory test that the uncertainties
associated with progenitor structure do not spoil the viability of the method. On the other hand, a firm conclusion on the other uncertainties requires further studies and goes beyond our present goals. In the following, however, we make some comments.

The EOS dependence explored in~\cite{Kachelriess:2004ds} at very early times seems to be quite moderate. We also argued in Sec.~\ref{qualdisc} that we do not expect significant changes within current uncertainties, at least for the early accretion phase. Additionally, the knowledge about the EOS has been progressing rapidly, both theoretically and observationally (see e.g.~\cite{Hebeler:2010jx} and refs. therein).  It is thus unclear to what extent the residual uncertainty in models compatible with all constraints {\it at the time of next SN detection} will hamper the viability of these diagnostics. Note that the neutrino lightcurve itself at later times may be used to constrain the equation of state (see e.g.~\cite{Roberts:2011yw} and refs. therein).

Concerning numerical model dependence, we had at our disposal also three simulations of the Basel/Darmstadt group (for details of their SN modeling, see~\cite{Liebendoerfer:2004}) for 10.8, 15 and 18 M$_\odot$ progenitors, using the physics input described in Ref.~\cite{Fischer:2009af}. A similar analysis as the one described in Sec. 4 shows that the physical effect  studied here ($\bar{\nu}_x$ signal rises faster than the $\bar{\nu}_e$ one) is always present, and differences between the NH and IH cases appear even more clearly, both due to more energetic $\bar{\nu}_x$ and more sudden rise of their luminosity.  However, we cannot compare these simulations directly with those of the Garching group since only the latter apply a treatment of weak  interactions including nucleon thermal motions, weak magnetism corrections, and energy transfer in nucleon recoils. On the other hand, tests performed by both groups demonstrate that inclusion of nucleon recoil effects lead to lower $\bar{\nu}_x$ energies as well as a slower luminosity rise, in the direction of reconciling the differences. Note that applying identical input physics to core-collapse SN simulations, both Basel/Darmstadt and Garching results agree qualitatively and quantitatively~\cite{Liebendoerfer:2005,Mueller:2010}.  For sure, it is known that an accurate treatment of issues like general relativity and neutrino-nucleon scatterings is needed to have reliable quantitative predictions of the observables: This is illustrated for example 
by Figs.~6 and 7 of Ref.~\cite{Mueller:2010}, as well as by Figs.~3 and 4 of the recent paper~\cite{Lentz:2011aa}.
 
For the time being,  we can tentatively conclude that the simulations of the Basel/Darmstadt group and the results presented in the recent paper~\cite{Lentz:2011aa}  confirm {\it qualitatively} the generality of the feature discussed here, but extra work will be required to establish wether  this agreement also holds quantitatively, namely if the mild spread due to progenitor structure found in the  simulations of the Garching group provides a reasonable estimate of the theoretical uncertainties. 

\section{Summary and conclusions}\label{conclusions}
We explored the chances of neutrino mass-hierarchy diagnostics by the $\overline{\nu}_e$ signals during  the (early) accretion phase of iron-core SNe, of the order of 100 ms or so. As already visible in some old simulations~\cite{Kachelriess:2004ds},  temporal profiles of $\bar{\nu}_x$ and $\bar{\nu}_e$ fluxes appear quite different in the accretion phase, and relatively model-independent: in particular,  the $\bar{\nu}_x$ signal rises faster than the $\bar{\nu}_e$ one. We have explored  the feasibility of using this observable for neutrino mass hierarchy determination in IceCube, which should provide an exquisite high-statistics determination of the lightcurve of a Galactic SN. 
For not too small values of $\theta_{13}$ favoured by recent data, the potentially ``clean'' theoretical interpretation of the measurement in terms of a given hierarchy is a consequence of the suppression of collective oscillations at early postbounce times by matter multi-angle effects and of the following evolution being dictated by MSW transitions. 

We have analyzed in detail the expected time profiles for ten different simulations carried out by the Garching group. We found that a {\it shape} difference in the rise time curve between the IH and NH cases is quite independent of the progenitor used (most notably of its mass); it does not appear that there is a major dependence from the dimensionality of the simulation either. We also showed with extensive Monte Carlo simulations that the expected statistical error is small if compared to the typical differences found between the two classes of models (NH vs. IH) and should not provide the limiting factor for diagnostics for typical distances, spectra and energetics  of a Galactic core-collapse SN.

While extra investigations are needed to assess quantitatively the ``theoretical errors'', i.e. the dependence of the signature discussed here on the equation of state, numerical schemes, etc. (see Sec.~\ref{discussion} for a brief discussion), in several other respects our analysis has been conservative. For example, diagnostic potential might be increased by  optimizing the statistical estimators, of which we introduced  here a few examples.
Further improvement in the diagnostic power may be obtained by  constraints on the spectral properties of the signal, either at existing Cherenkov detectors like Super-Kamiokande or even via modest spectral sensitivity at IceCube or (more likely) at its upgrades~\cite{Demiroers:2011am}: models inconsistent with the measured spectrum might be excluded and reduce the theoretical error.
Also, other complementary techniques have been proposed to extract information on the mass hierarchy from SN neutrino signals, the most notable one being through spectral modulations induced via Earth matter effect (see, e.g., Ref.~\cite{Dighe:2003be}).  The interesting point is that, different from the Earth effect, the signature discussed here appears to be independent of a serendipitous position of the detector at the arrival time of the SN signal. We checked in fact that the signal is only altered on a negligible level of percent when the SN is observed throughthe Earth. Also, differently from  shock effects, it is independent of poorly understood details of the flavor evolution in the SN during the complex cooling phase. On the other hand, {\it if observed}, signatures like the Earth matter effect are less worrisome in terms of model dependence. 
Additionally, if a complementary detector of $\nu_e$'s was available~\cite{GilBotella:2003sz}, one might hope to cross-check the preferred solution (IH or NH) from rise time information with the inference on the hierarchy from the detection or absence of the neutronization burst~\cite{Dighe:1999bi}.  In general, any Megaton class detector with flavor and energy resolution  will of course improve the chances of both astrophysics and physics diagnostics. 
It is also expected that when the next galactic SN will eventually occur, the simulations will be calibrated to reproduce that particular event: Informations on the progenitor should allow a reduction in the theoretical uncertainties.

The arguments developed in this article confirm once more the high physics potential of SN neutrinos in shedding light on the still unknown pieces of the neutrino mass and mixing framework. We hope that the potential importance
for physical diagnostics of the signature discussed here will motivate further attention and progress in numerical simulations. Therefore, even though a galactic SN explosion is a rare event, we are sure that the patient waiting will eventually be rewarded with a bonanza of information.
\section*{Acknowledgments} 

We thank Ricard Tom{\`a}s for collaboration during the initial development of this project.
We also thank  Eligio Lisi, Antonio Marrone, Georg Raffelt, Irene Tamborra for reading the manuscript and for useful comments on it. 
L.H.\ and H.-T.J.\ are grateful to Andreas Marek and Bernhard M\"uller for their support.
The work of S.C., A.M. was supported by the German Science Foundation (DFG) within the Collaborative Research Center 676 ``Particles, Strings and the Early Universe''. 
T.F. acknowledges support from the Swiss National Science Foundation (SNF) under grant~no.~PBBSP2-133378 and HIC for FAIR. At Garching, the project was partially  funded by the Deutsche Forschungsgemeinschaft through the Transregional Collaborative Research Centers SFB/TR~27 ``Neutrinos and Beyond'' and SFB/TR~7 ``Gravitational Wave Astronomy'', and the Cluster of Excellence EXC~153 ``Origin and Structure of the Universe'' ({\tt http://www.universe-cluster.de}). The supernova simulations were possible by computer time grants at the John von Neumann Institute for Computing (NIC) in J\"ulich, the H\"ochst\-leistungs\-re\-chen\-zentrum of the Stuttgart University (HLRS) under grant number SuperN/12758, the Leib\-niz-Re\-chen\-zentrum M\"unchen, and the RZG in Garching.


\end{document}